\documentclass[preprint,showpacs,aps,fleqn]{revtex4}%
\usepackage{graphicx}
\usepackage{psfrag}
\usepackage{amsmath}
\usepackage{subfigure}

\newcommand{\be}{\begin{equation}}
\newcommand{\ee}{\end{equation}}
\newcommand{\bea}{\begin{eqnarray}}
\newcommand{\eea}{\end{eqnarray}}
\renewcommand{\vec}[1]{\boldsymbol{#1}}

\begin{document}

\title{Classical Strongly Coupled QGP: \\
VII. Energy Loss}

\author{Sungtae Cho}
\email{scho@grad.physics.sunysb.edu}
\author{Ismail Zahed}
\email{zahed@zahed.physics.sunysb.edu}
\address{Department of Physics and Astronomy \\
State University of New York, Stony Brook, NY, 11794}

\begin{abstract}
We use linear response analysis and
the fluctuation-dissipation theorem to derive
the energy loss of a heavy quark in the SU(2) classical
Coulomb plasma in terms of the $l=1$
monopole and non-static structure factor. The result is
valid for all Coulomb couplings $\Gamma=V/K$, the
ratio of the mean potential to kinetic energy.  We use
the Liouville equation in the collisionless limit to assess
the SU(2) non-static structure factor. We find the energy
loss to be strongly dependent on $\Gamma$. In the
liquid phase with $\Gamma\approx 4$, the energy loss
is mostly metallic and soundless with neither a Cerenkov
 nor a Mach cone. Our analytical results compare
 favorably with the SU(2) molecular dynamics simulations
 at large momentum and for heavy quark masses.
   \end{abstract}


\maketitle

\newpage

\section{Introduction}
Parton energy loss at RHIC is widely viewed as a way to probe the
properties of the medium created during the first few fm/c of the
collision.  The medium is suspected to be a strongly coupled
liquid~\cite{shuryak2} with near perfect fluidity and strong
energy loss.

There have been a number of calculations involving parton
collisional~\cite{thoma&gyulassy, braaten&thoma, braaten&thoma2,
djordjevic} and radiative~\cite{djordjevic&heinz,
djordjevic&heinz2} energy loss at RHIC with the chief consequence
of jet quenching \cite{dumitruetal}. The measured jet quenching at
RHIC exceeds most current theoretical predictions, most of which
are based on a weakly coupled quark-gluon plasma (wQGP).

The QCD matter probed numerically using lattice simulations and at
RHIC using heavy ion collisions, is likely to be dominated by
temperatures in the few $T_c$ range making it de facto
non-perturbative. Non-perturbative methods are therefore welcome
for analyzing the QCD matter conditions in this temperature range.
An example being the holographic method as a tool for jet
quenching analysis~\cite{sin&zahed,liuetal}.

In this letter, we follow the approach suggested
in~\cite{gelmanetal, gelmanetal2, cho&zahed, cho&zahed2} to model
the strongly coupled quark and gluon plasma, by classical colored
constituents interacting via strong Coulomb interactions. This
 model has been initially analyzed using Molecular Dynamics (MD)
 simulations mostly for the SU(2) version with species of constituents
 (gluons).  The MD results reveal a strongly coupled liquid at
 $\Gamma\approx 4$ the ratio of the mean kinetic to Coulomb energy
 (modulo statistical fluctuations).
 The fractional energy loss is also found to be considerably
 larger than most leading order QCD estimates.

 Here, we will
 provide the analytical framework to analyze the MD simulation
 results for partonic energy loss in the cQGP.  In section 2, we
 outline a formal derivation of the energy loss in the cQGP for arbitrary
 values of $\Gamma$. In section 3, we use linear response theory
 and the fluctuation-dissipation theorem  to tie the energy loss to
 the non-static colored structure factor. In section 4, we derive
 explicitly the non-static structure factor using the Liouville equation.
Some useful aspects of the plasmon excitations in the cQGP are
discussed in section 5. In section 6, we analyze the energy loss
for both charm and bottom for $\Gamma=$2,3 and 4 in the liquid
phase and compare them to the recent SU(2) MD simulations
\cite{dusling&zahed}. In section 7, we discuss the relevance of
our results to RHIC and holographic QCD.

\section{Energy Loss}

\renewcommand{\theequation}{II.\arabic{equation}}
\setcounter{equation}{0}

Consider an SU(2) colored particle of charge $q^a$ travelling with velocity $v$
in the strongly coupled colored plasma~\cite{gelmanetal}. The equation of motion
of this {\it extra} particle in phase space follows from the Poisson bracket

\begin{equation}
\frac{d\vec p_i}{dt}=-\{H,\vec p_i\}=q^a\cdot \vec E_{in}^a
\label{eq001p}
\end{equation}
with the longitudinal colored electric field

\begin{equation}
\vec E_{in}^a=-\vec \nabla\sum_{i}\frac{Q^a_i(t)}{|\vec r-\vec
r_i(t)|}=-\vec \nabla_i\Phi^a_{in}(t,\vec r) \label{eq002p}
\end{equation}
We note that in~\cite{gelmanetal} the SU(2) plasma is considered
mostly electric with massive constituents $m\beta\approx 3$.  As a
result the transverse electric contribution is absent
in~(\ref{eq002p}). Also, (\ref{eq001p}) does not involve the
magnetic part  of the Lorentz force for the same reasons.  The
latter is irrelevant for the energy loss per travel length $\vec
r=\vec vt$

\begin{equation}
\frac{dK}{dr}=\vec v q^a\vec E_{in}^a(t,\vec r=\vec v t)
\label{eq003p}
\end{equation}
even in the ultrarelativistic case since the magnetic force does not perform work.

The induced colored Coulomb potential $\Phi_{ind}$ follows from the total
colored potential $\Phi_{\rm tot}$ through

\begin{equation}
\Phi^a_{tot}(\omega,\vec k)=\Phi^a_{ind}(\omega,\vec
k)+\Phi^a_{ex}(\omega,\vec k)=\frac{\Phi^a_{ex}(\omega,\vec
k)}{\epsilon_L(\omega,\vec k)} \label{eq004p}
\end{equation}
The last relation defines the longitudinal dielectric constant
with $\vec \Phi_{ex}(\omega,\vec k)=\frac{4\pi}{k^2}2\pi\vec q
\delta(\omega-\vec k\cdot\vec v)$,  the colored potential caused
by the {\it extra} particle in the probe approximation (ignoring
back reaction). Thus

\begin{equation}
 \Phi^a_{ind}(t,\vec r)=q^a\int \frac{d\vec k}{(2\pi)^3}
\left(\frac{1}{\epsilon_L(k\cdot v, \vec k)}-1\right)\,
\frac{4\pi}{k^2}e^{i\vec k\cdot\vec r-i\vec k\cdot\vec v t}
\label{eq005p}
\end{equation}
Using (\ref{eq002p}) and (\ref{eq003p}) we have for the energy
loss of a fast moving probe SU(2) charge

\begin{equation}
-\frac{dK}{dr}=-\frac{\vec q^2}{\pi v^2}\int
\frac{dk}{k}\int_{-kv}^{kv}\omega d\omega \Im\left(\frac{1}{
\epsilon_{L}(\omega+i0,\vec k)}\right)
\label{eq006p}
\end{equation}
after using the analytical property of
$\epsilon_L(z,k)=\epsilon_L(-z^*,-k)$ which follows from the
causal character of the longitudinal dielectric function as
detailed below.  (\ref{eq006p}) is identical in form to the one
derived for the Abelian one component colored Coulomb plasma
in~\cite{ichimaru}, to the exception of the SU(2) classical
Casimir $\vec{q}^2$ in (\ref{eq006p}). It is different in content
through the longitudinal dielectric constant $\epsilon_L$ which
now should be derived for a colored SU(2) Coulomb plasma. Our
derivation is {\it fully} non-Abelian in the probe approximation.

\begin{figure}[!h]
\begin{center}
\includegraphics[width=0.55\textwidth]{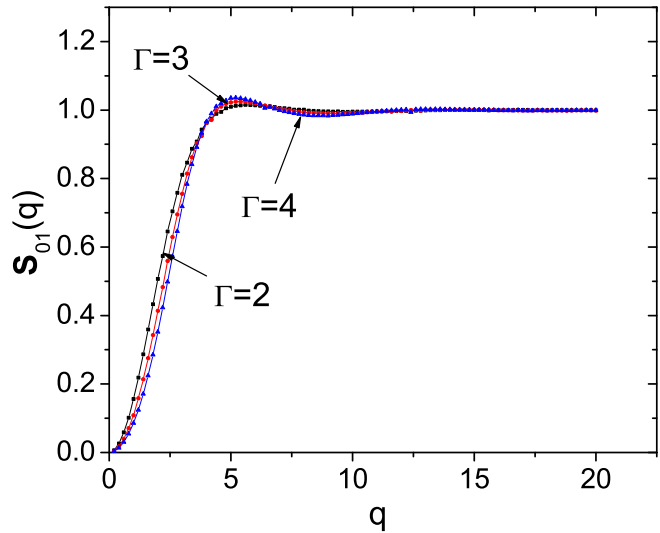}
\end{center}
\caption{Static structure factors for $\Gamma=2,3,4$} \label{structure_G020304}
\label{S011}
\end{figure}

\begin{figure}[!h]
\begin{center}
\includegraphics[width=0.49\textwidth]{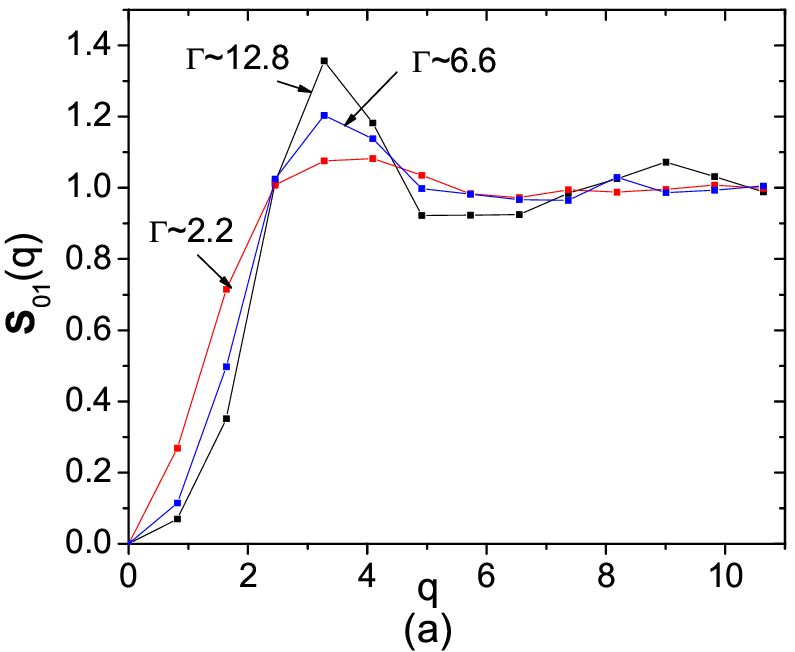}
\includegraphics[width=0.49\textwidth]{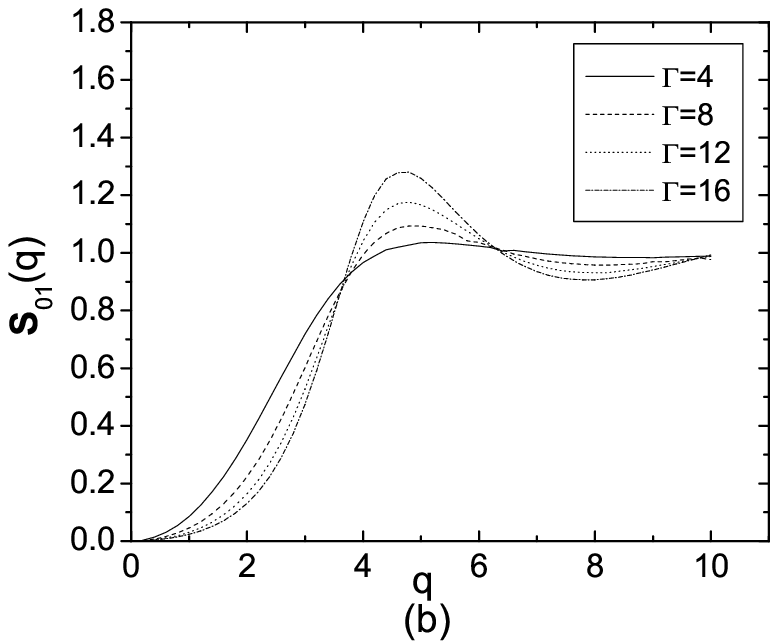}
\end{center}
\caption{${\bf S}_{01}(q)$:
molecular dynamics simulation (a) and analytic (b). See text.}
\label{structure}
\end{figure}

Below we show that for the SU(2) colored Coulomb plasma at strong Coulomb
coupling, (\ref{eq006p}) reads

\begin{equation}
-\frac{dK}{dr}=-\frac{\vec q^2}{\pi v^2}\int \,
\frac{dk}{k^3}\,\frac{k_D^2\,{\bf S}_{01}(k)}{1-{\bf S}_{01}(k)}
\int_{-kv}^{+kv}\, d\omega\,\omega\,\Im\left(\frac{1}{
\epsilon_{1}(\omega+i0,\vec k)}\right)
\label{LIN12}
\end{equation}
with $k_D^2$ the SU(2) Debye wave number squared. Here

\begin{eqnarray}
\epsilon_1(z,\vec k)=1-n\,{\bf c}_{D1}(\vec k)\,{\bf W}(z/\omega_T)
\label{LIN12X}
\end{eqnarray}
with the thermal frequency $\omega_T=v_Tk$ and velocity
$v_T=\sqrt{T/m}$ and

\begin{equation}
{\bf W}(z/\omega_T)=\frac{1}{\sqrt{2\pi}}\int_{-\infty}^{+\infty}dt\,\frac{t}{t-z/\omega_T}\,{e^{-t^2/2}}
\label{LIN13}
\end{equation}
The l=1 static structure factor ${\bf S}_{01}$~\cite{cho&zahed3}

\begin{equation}
{\bf S}_{01}(k)=\left<\left|\sum^N_{j=1}e^{i\vec k\cdot \vec
r_j}\,Q_j^a\right|^2\right> \label{S01}
\end{equation} satisfies the
generalized Ornstein-Zernicke equation

\begin{equation}
{\bf S}_{01}(k)=\frac 1{1-n{\bf c}_{D1}(k)}
\end{equation}
in the colored Coulomb plasma with 1 species density $n=N/V$.
In Fig.~\ref{S011} we show
analytical results for (\ref{S01}) around the liquid point~\cite{cho&zahed3}.
In Fig.~\ref{structure}a  we show the behavior of (\ref{S01}) using
SU(2) molecular dynamics simulations with the dimensionless
wavenumber $q=k\,a_{WS}$ where $a_{WS}$ is the Wigner-Seitz
radius through $1/n=4\pi\,a_{WS}^3/3$. In Fig.~\ref{structure}b we show the
analytical results for the same range of $\Gamma$ in~\cite{cho&zahed3}.

The
$l=1$ contribution $\epsilon_1$ plays the role of a generalized
longitudinal dielectric constant in the SU(2) Coulomb plasma.
Indeed, for weak Coulomb coupling $\Gamma\ll 1$, $-n{\bf
c}_{D1}\approx k_D^2/k^2$ so that ${\bf S}_{01}\approx
k^2/(k^2+k_D^2)$. The energy loss (\ref{LIN12}) reduces to
(\ref{eq006p}) with $\epsilon_L\rightarrow \epsilon_1$. At weak
coupling $\epsilon_1$ in (\ref{LIN12X}) is the standard Vlasov
dielectric function in~\cite{ichimaru}. The only difference is in
the SU(2) Debye wave number.

\section{Linear Response}

\renewcommand{\theequation}{III.\arabic{equation}}
\setcounter{equation}{0}

To construc the longitudinal dielectric constant for the SU(2)
Coulomb plasma we will make use of the Liouville kinetic equations
for the time dependent structure factors derived
in~\cite{cho&zahed3}. For that we recall that in linear response,
the induced color charge density $\rho_{ind}^a=\nabla\cdot
E_{ind}^a/4\pi$  ties with the external potential $\Phi_{ext}^b$
through the retarded correlator

\begin{equation}
\rho_{ind}^a(t,\vec r)=i\int\,dt'\,d\vec r'\, \left<{\bf
R}\left({\bf J}_0^a(t,\vec r){\bf J}_0^b(t',\vec
r')\right)\right>\,{\Phi}_{ext}^b (t',\vec r') \label{LIN1}
\end{equation}
where ${\bf J}_0^a$ are the pertinent color charge densities.  In Fourier space
we have

\begin{equation}
\Phi_{ind}^a=-\frac{4\pi}{k^2}\,\Delta^{ab}_R(\omega,\vec
k)\,\Phi_{ext}^b \label{LIN2}
\end{equation}with

\begin{equation}
\Delta_R^{ab}(\omega, \vec k) =-i\int \,e^{-i\omega t+i\vec k\cdot
\vec r}\, \left<{\bf R}\left({\bf J}_0^a(t,\vec r){\bf
J}_0^b(t',\vec r')\right)\right> \label{LIN3}
\end{equation}
A comparison
of (\ref{LIN2}) with (\ref{eq004p}) yields

\begin{equation}
\left(\frac 1{\epsilon_L(\omega, \vec
k)}-1\right)\,\delta^{ab}=-\frac{4\pi}{k^2}\,\Delta_R^{ab}(\omega,
\vec k) \label{LIN5X}
\end{equation}
which defines the longitudinal dielectric constant.

The retarded correlator (\ref{LIN3}) is in general a quantum object, we now show
how to extract it from the correlations in the classical and strongly coupled SU(2)
colored Coulomb plasma. For that, we note that the colored charge density in the SU(2) phase space is

\begin{equation}
{\bf J}_0^a(t,\vec{r})=\int\,dQ\,d\vec p\, Q^a\delta f(t,\vec r,
\vec p, \vec Q ) \label{CHARGE1}
\end{equation}
and that the SU(2)  charge-charge correlator is

\begin{equation}
\left<{\bf J}_0^a(t,\vec{r}){\bf J}_0^b(t',\vec{r}')\right> =\frac
13 \delta^{ab} \int\,dQ\,dQ'\,d\vec p\,d\vec p'\,\vec Q\cdot \vec
Q'\, {\bf S}(t-t', \vec r-\vec r', \vec p\vec p' ,\vec Q\cdot \vec
Q') \label{CHARGE2}
\end{equation}
where global time, space and color invariances were used thanks to
the statistical averaging. The time dependent structure factor
${\bf S}=\langle\delta f\delta f\rangle$ was defined
in~\cite{cho&zahed3} . Using the color Legendre transform of ${\bf
S}$ yields

\begin{equation}
\left<{\bf J}_0^a(t,\vec{r}){\bf J}_0^b(t',\vec{r}')\right>
=\delta^{ab} \int\,d\vec p\,d\vec p'\,
{\bf S}_1(t-t', \vec r-\vec r', \vec p\vec p')
\label{CHARGE3}
\end{equation}
Only the $l=1$ partial wave in the Legendre transform of the color part of
${\bf S}$ contributes to the SU(2) charge-charge correlation function.

The fluctuation-dissipation theorem in the {\it classical limit} ties the retarded
correlator $\Delta_R$  in (\ref{LIN3}) to the Fourier transform of the
classical phase space fluctuations (\ref{CHARGE3}) as

\begin{equation}
\Im\Delta^{ab}_R(\omega, \vec
k)=\delta^{ab}\frac{n\omega}{2T}\,{\bf S}_1(\omega, \vec k)\equiv
-\delta^{ab}\frac{n\omega}{T}\,\Im{\bf S}_1(z, \vec k)
\label{LIN5}
\end{equation}
The last relation follows from ${\bf S}_1(\omega, \vec k)=-2\,{\rm
Im}\,{\bf S}_1(\omega, \vec k)$ between the Laplace transform and
Fourier transform of ${\bf S}_1$ with $z=\omega+i0$.

\section{Non-Static Structure Factor}

\renewcommand{\theequation}{IV.\arabic{equation}}
\setcounter{equation}{0}

We have shown in~\cite{cho&zahed4} that the l-color partial wave
of the Laplace transform of ${\bf S}_l$ obeys the Liouville
equation

\begin{equation}
z{\bf S}_l(z\vec k ;\vec p \vec p')-\int d\vec p_1 \Sigma_l(z\vec k
;\vec p \vec p_1)\,{\bf S}_l(z\vec k ;\vec p_1 \vec p')={\bf
S}_{0l}(\vec k ;\vec p \vec p') \label{eq007e}
\end{equation}
${\bf S}_{0l}$ is the $l$ static structure factor introduced
in~\cite{cho&zahed3}

\begin{equation}
{\bf S}_{0l}(\vec k ;\vec p \vec p')=n\,f_0(\vec p)\,\delta(\vec
p-\vec p')+n^2 f_0(\vec p)\,f_0(\vec p')\,{\bf h}_l(\vec k)
\label{LIN7}
\end{equation}
with the Maxwell-Boltzmann distribution $f_0(\vec p)$. The
structure factor ${\bf h}_l(\vec k)$ relates to the standard
structure factor ${\bf S}_{0l}(k)$ by the generalized
Ornstein-Zernicke equations

\begin{equation}
\frac 1n \,\int \,d\vec{p}\,d\vec p' \,{\bf S}_{0l}(k ;\vec p \vec
p') ={\bf S}_{0l}(k)=1+n\,{\bf h}_l(k)=\left(1-n\,{\bf
c}_{Dl}(k)\right)^{-1} \label{LIN8}
\end{equation}
The self-energy kernel $\Sigma_l$ in (\ref{eq007e}) splits into a
static and collisional contribution in each color partial wave $l$
\cite{cho&zahed4}.

We note that

\begin{equation}
{\bf S}_l(zk)=\frac 1n\,\int d\vec{p}\,d\vec p' \,{\bf S}_l(z\vec
k ;\vec p \vec p') \label{LIN6}
\end{equation}
with $l=1$ is what is needed in (\ref{LIN5}). For that, we solve
(\ref{eq007e}) in the collisionless limit with

\begin{equation}
\Sigma_l(z \vec k ;\vec p \vec p')\approx \frac{1}{m}\vec
k\cdot\vec p\,\delta(\vec p-\vec p')-\frac{1}{m}\vec k\cdot\vec
p\,n\,f_0(\vec p)\,{\bf c}_{Dl}(\vec k) \label{LIN7}
\end{equation}
We recall that the SU(2)  color part of the Liouville operator is
a genuine 3-body force that only enters the collisional
contribution.~\cite{cho&zahed4}. Inserting (\ref{LIN7}) into
(\ref{eq007e}) and using (\ref{LIN7}) and (\ref{LIN8}) yield in
the collisionless limit

\begin{equation}
{\bf S}_l(z,\vec k)=\frac{{\bf S}_{0l}(k)}{\epsilon_l(z,\vec
k)}\,\int\, d\vec{p}\,\frac{f_0(p)}{z-\vec k\cdot \vec p/m}
\label{LIN9}
\end{equation}
with

\begin{equation}
\epsilon_l (z,\vec k)=1+n\,{\bf c}_{Dl} (k)\int\,d\vec p\,
\frac{\vec k\cdot \vec p/m}{z-\vec k\cdot\vec p/m}\,f_0(p)
\label{LIN10}
\end{equation}
and

\begin{equation}
\int\, d\vec{p}\,\frac{f_0(p)}{z-\vec k\cdot\vec
p/m}=\frac{1}{\omega}\Big(1-{\bf W}(z/\omega_T)\Big)
\end{equation}
If we insert (\ref{LIN9}) into (\ref{LIN5}) and then use
(\ref{LIN5X}), we find for $l=1$

\begin{equation}
\Im\,\frac 1{\epsilon_L(z, \vec
k)}=-\frac{k_D^2}{k^2}\,\frac{1}{n{\bf c}_{D1}(k)} \,\,\Im\,\frac
1{\epsilon_1(z, \vec k)} \label{LIN11}
\end{equation}
Inserting (\ref{LIN11}) into (\ref{eq006p}) yields the announced
relation (\ref{LIN12}).

\section{SU(2) Plasmon}

\renewcommand{\theequation}{V.\arabic{equation}}
\setcounter{equation}{0}

Before analyzing the energy loss in (\ref{LIN12X}) for heavy charged probes, it is instructive to discuss
the zeros of the longitudinal dielectric constant $\epsilon_1(\omega, k)=0$ in (\ref{LIN12X}) as they reflect
on the longitudinal excitations in the $l=1$ channel. For that, we need the behavior
of ${\bf W}(x)$ as defined in (\ref{LIN13}) with $x=\omega/v_Tk$ for small and large ratio $k/k_D$.
$v_T=\sqrt{T/m}$ is the velocity of the the particles in the SU(2) heat bath. In weak coupling QCD
$m\approx gT$, while in strong coupling $m\approx \pi T$.

In general,

\begin{equation}
{\bf W}(x)={\bf W}_R(x)+i{\bf
W}_I(x)=1-xe^{-x^2/2}\,\psi(x)+i\sqrt{\pi\over 2}\,x\,e^{-x^2/2}
\label{W1}
\end{equation}
with $\psi(x)=\int_0^x\,dy\,e^{y^2/2}$ the incomplete exponential
function. For $k\ll k_D$ or $x\gg 1$,

\begin{equation}
{\bf W}(x)\approx -\frac{1}{x^2}+i{\sqrt{\pi}\over 2} \,xe\,^{-x^2}
\label{LIN14}
\end{equation}
while for $k\gg k_D$ or $x\ll 1$

\begin{equation}
{\bf W}(x)\approx 1-x^2+i{\sqrt{\pi}\over 2} \,xe\,^{-x^2}
\label{LIN14X}
\end{equation}
So in the long wavelength limit with $k\ll k_D$,
(\ref{LIN12X}) expands to

\begin{equation}
\epsilon_1(\omega,\vec k)\approx 1+\frac{n{\bf
c}_{D1}(k)}{x^2}\left(1-i{\sqrt{\pi}\over 2}x^3e^{-x^2/2}\right)
\label{LIN15}
\end{equation}
For small $k$, $n{\bf c}_{D1}(k)\approx {\bf S}_{01}(k)\approx
k_D^2/k^2$ whatever the coupling in the SU(2) colored plasma. Thus

\begin{equation}
\epsilon_1(\omega,\vec k)\approx
1-\frac{\omega_p^2}{\omega^2}\left(1-i{\sqrt{\pi}\over
2}x^3e^{-x^2/2}\right) \label{LIN16}
\end{equation}
with the plasmon frequency $\omega_p=v_T\,k_D$.  So for $k\ll
k_D$, the zero of (\ref{LIN16}) is

\begin{equation}
\omega^2_1(k)\approx\omega_p^2\,\left(1-i{\sqrt{\pi}\over
2}\,\frac{k_D^3}{k^3}e^{-k_D^2/2k^2}\right) \label{LIN17}
\end{equation}
The SU(2) colored Coulomb plasma supports a plasmon with frequency
$\omega_p$ with an exponentially small width
$e^{-\omega^2/2v_T^2k^2}$ both at weak and strong SU(2) Coulomb
coupling $\Gamma$.  This result agrees with our analytic and
leading kinetic analysis in the hydrodynamical
limit~\cite{cho&zahed4} . The current analysis provides the
non-analytic imaginary part as well.

The high $k\gg k_D$ limit is metallic whatever $\Gamma$ with

\begin{equation}
\epsilon_1(x)\approx 1-in{\bf c}_{D1}(k)\,x{\sqrt{\pi\over
2}}\,e^{-x^2/2} \label{LIN17X}
\end{equation}
with a metallic conductivity $x=\omega/\omega_T=\omega/v_Tk$

\begin{equation}
\sigma_1(\omega, \vec k)=\frac{n\,{\bf
c}_{D1}(k)}{\sqrt{32\pi}}\,{{\omega^2}\over\,v_Tk}\,e^{-\omega^2/2v_T^2k^2}
\label{META}
\end{equation}
We note that the plasmon branch disappears at high $k$ in
(\ref{LIN17X})  as the plasma turns metallic i.e. a collection of
free colored SU(2) particles with a classical thermal spectrum.
Also the plasmon in (\ref{LIN16}) broadens substantially at
$k\approx k_D$ with its real part comparable to its imaginary
part. This point causes the plasmon contribution to drop from the
energy loss in the colored SU(2) Coulomb plasma as we show below.

\begin{figure}[!h]
\begin{center}
\includegraphics[width=0.50\textwidth]{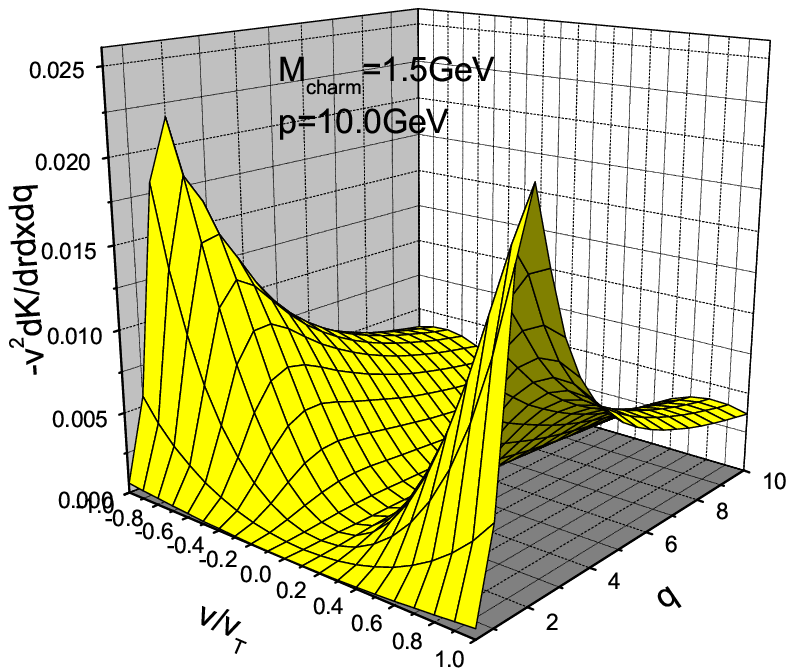}
\includegraphics[width=0.49\textwidth]{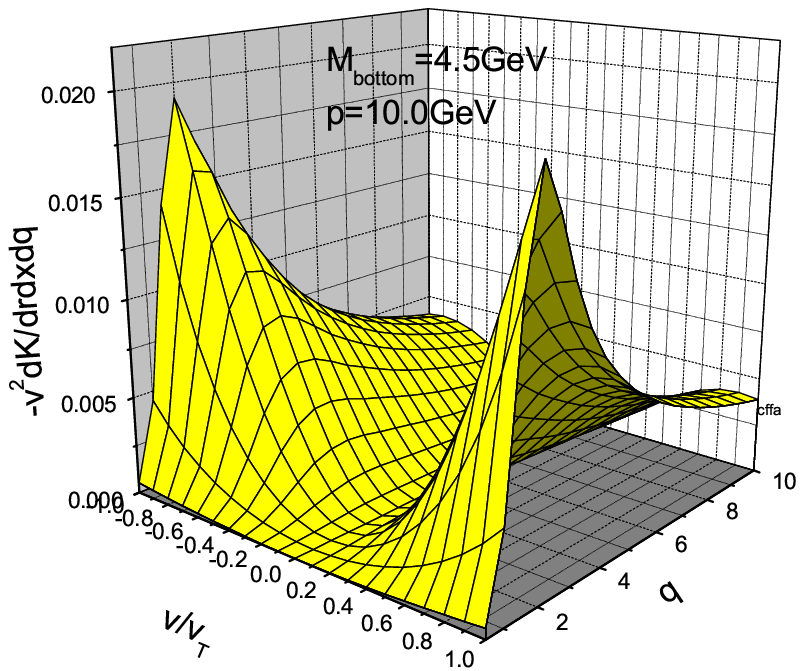}
\caption{Surface plot of $-v^2 dK/drdxdq$ for charm and bottom. See text.} \label{3dplot}
\end{center}
\begin{center}
\includegraphics[width=0.50\textwidth]{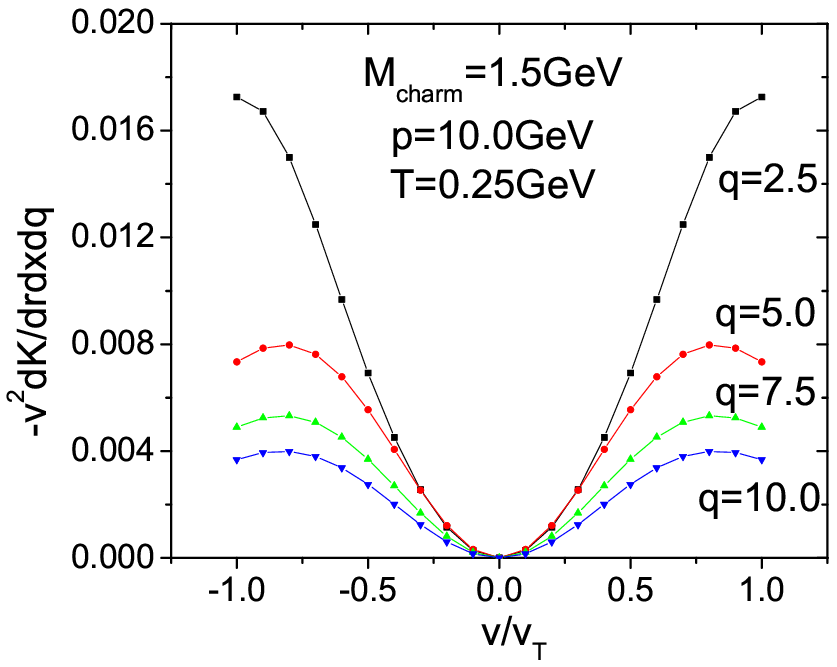}
\includegraphics[width=0.49\textwidth]{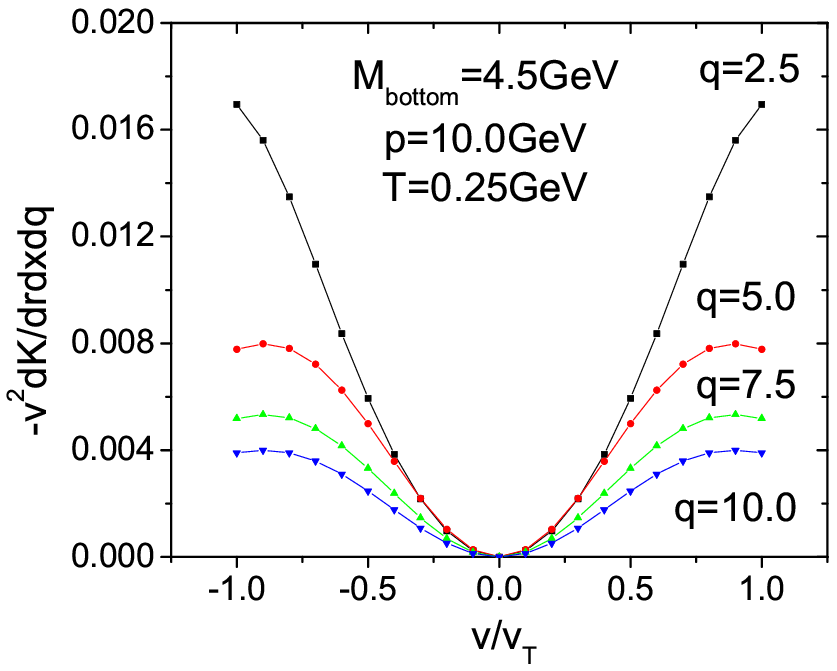}
\end{center}
\caption{$-v^2 dK/drdxdq$ versus $v/v_T$ for charm and bottom
quark for fixed $q$. See text.} \label{2dplot}
\end{figure}

\section{Charm and Bottom Loss}

\renewcommand{\theequation}{VI.\arabic{equation}}
\setcounter{equation}{0}

Inserting (\ref{LIN12X}) into (\ref{LIN12}) and using the explicit form (\ref{W1}) yields
the energy loss in the SU(2) Coulomb plasma

\begin{eqnarray}
-\frac{dK}{dr}=&&\frac{g^2\,C_F}{4\pi}\frac{\omega_p^2}{v^2}\int_0^{k_{max}}\,{dk}\,\frac{1}{k}\,\nonumber\\
&&\times\frac{1}{\sqrt{2\pi}}\int_{-v/v_T}^{v/v_T}\,dx\,e^{x^2/2}\,
\Bigg(\bigg((1-n{\bf c}_{D1}(k))\,e^{x^2/2}/x+n{\bf c}_{D1}(k)\,\psi(x)\bigg)^2+\pi\,n^2{\bf c}^2_{D1}(k)/2\Bigg)^{-1}\nonumber\\
\label{CB1}
\end{eqnarray}
For an SU(2) probe charge after the substitution $\vec{q}^2\rightarrow g^2C_F/4\pi$ with $C_F$ the SU(2) Casimir.
We note that (\ref{CB1}) is cutoff in the infrared by the Debye wave number since ${\bf S}_{01}(k)\approx k^2/k_D^2$.
So the main contribution to the energy loss in (\ref{CB1}) stems from the region $k>k_D$ for which the SU(2) plasmon
is too broad to contribute as we noted earlier.  Most of the loss stems from the metallic part of the SU(2) plasma which
is the analogue as rescattering against the {\it free} thermal spectrum explicit in (\ref{META}).

In Fig.~\ref{3dplot} we display the integrand in (\ref{CB1}) versus the jet velocity $v/v_T$ and
the dimensionless momentum $q=ka_{WS}$.  This is a weighted plot of the longitudinal spectral function along the
jet velocity. The two wings at small $q$ are the two plasmons peaks, which progressively
turns into the thermal distribution at larger $q$.  In Fig.~\ref{2dplot} we show the same integrand
for fixed $q$ versus the jet velocity $v/v_T$ normalized to the thermal velocity $V_T$.  We note
again the 2 plasmon poles around $v\approx v_T$ at small $q$. The vanishing of the termal
distribution at  $q=0$ follows from the extra $x^2$ weight arising from the denominator of
(\ref{CB1}) for ${\bf c}_{D1}(k)\approx 0$ at large $k$.

Since the loss is colored with only $l=1$ contributing  and is
metallic with only $k>k_D$ contributing,  we do not see colored
Cherenkov radiation stemming from plasmon
emission~\cite{ruppert&muller}, nor the ubiquitous Mach cone
stemming from coupling to the sound mode~\cite{casalderryetal}.
While the sound mode contributes to ${\bf S}$ in (\ref{CHARGE2})
it drops in the statistical averaging as only $l=1$ or plasmon
channel contributes. The energy loss in the classical colored
SU(2) Coulomb plasma is mostly metallic with $k>k_D$ and soundless
due to the  color quantum numbers of the fast moving  probe
charge.

A qualitative estimate for the energy loss follows by using ${\bf S}_{01}(k)\approx k^2/(k^2+k_D^2)$ and saturating
the integrand by $k>k_D$,

\begin{equation}
-\frac{dK}{dr}\approx \frac{g^2\,C_F}{4\pi}\frac{\omega_p^2}{v^2}
\,\left(\sqrt{2\over \pi}\int_{0}^{v/v_T}\,x^2\,e^{-x^2/2}
\right)\,\,\ln{\left(\frac{k_{max}}{k_D}\right)} \label{CB2}
\end{equation}
The upper divergence is manifest in (\ref{CB1}) at $k\gg k_D$
since ${\bf S}_{01}(k)\approx 1$ and ${\bf c}_{D1}(k)\approx
k_D^2/k^2$ through the generalized Ornstein-Zernicke equation for
all Coulomb couplings. The upper cutoff $k_{max}\approx 2\gamma \,
mv$ which is set by the maximum momentum transfer to the thermal
particle of mass $m$ in the rest frame of the probe particle $M\gg
m$. Typically $M$ is charm and bottom, while $m\approx gT$ in weak
coupling and $m\approx \pi T$ in strong coupling for a QCD plasma
near the critical point.  For the former $v/v_T\approx v\sqrt{g}$
(weak coupling) while for the latter $v/v_T\approx v\sqrt{\pi}$
(strong coupling).  For $v/v_T\gg 1$ (\ref{CB2}) reduces further
to

\begin{equation}
 -\frac{dK}{dx}\approx \frac{g^2\,C_F}{4\pi}\frac{\omega_p^2}{v^2}\,\,\ln{\left(\frac{2\gamma
 \,m\,v}{k_D}\right)}
\label{CB3}
\end{equation}
For the SU(2) colored Coulomb plasma. Aside from the Casimirs,
this result is analogous to the energy loss in the classical and
Abelian Coulomb plasma \cite{ichimaru, thoma}.

To assess the energy loss for varying Coulomb coupling
$\Gamma=({g^2C_2}/{4\pi})({\beta}/{a_{WS}})$, we will
rewrite the energy loss ~(\ref{CB1}) as

\begin{figure}[!h]
\begin{center}
\includegraphics[width=0.49\textwidth]{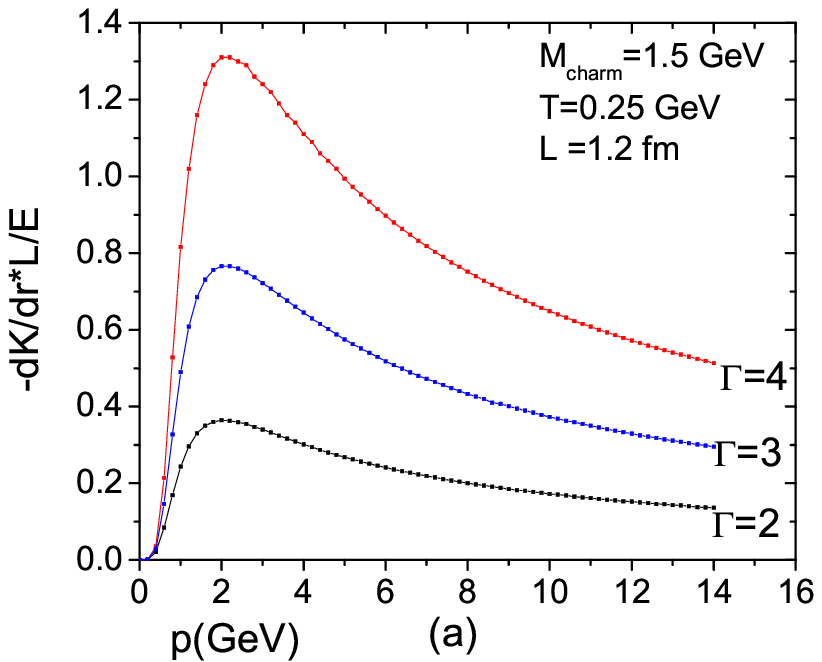}
\includegraphics[width=0.49\textwidth]{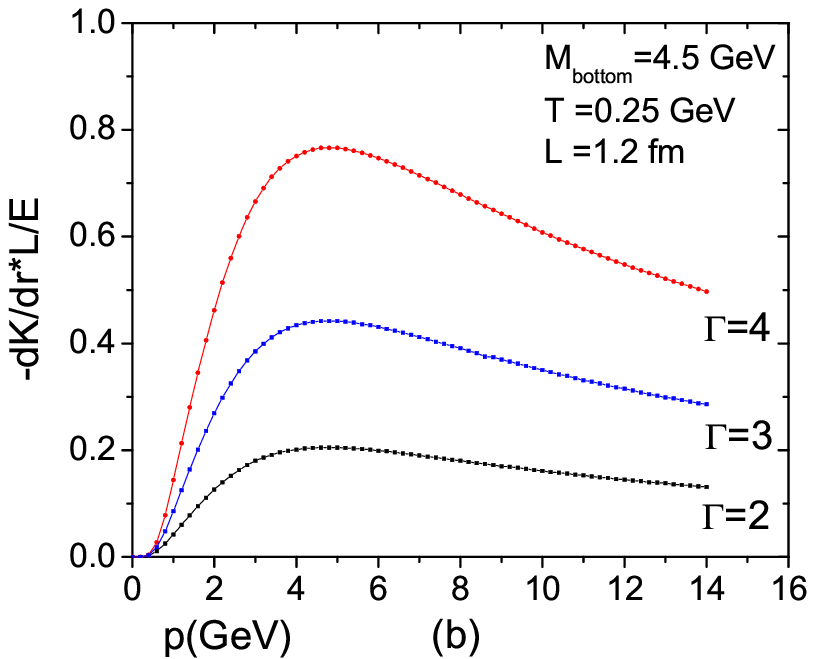}
\end{center}
\caption{Energy loss for charm (left) and bottom (right) in the cQGP: $\Gamma=2,3,4$.}
\label{charm_bottom1}
\end{figure}

\begin{figure}[!h]
\begin{center}
\includegraphics[width=0.49\textwidth]{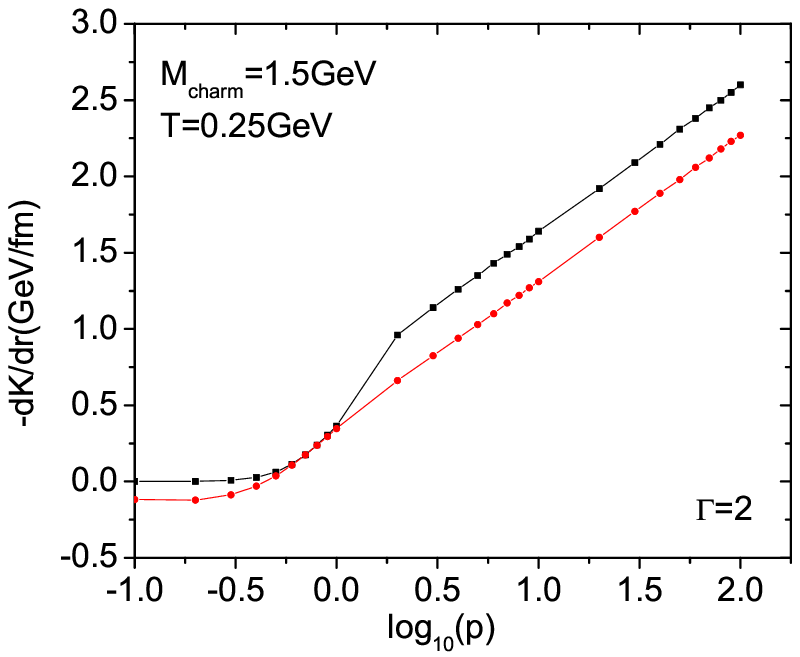}
\includegraphics[width=0.49\textwidth]{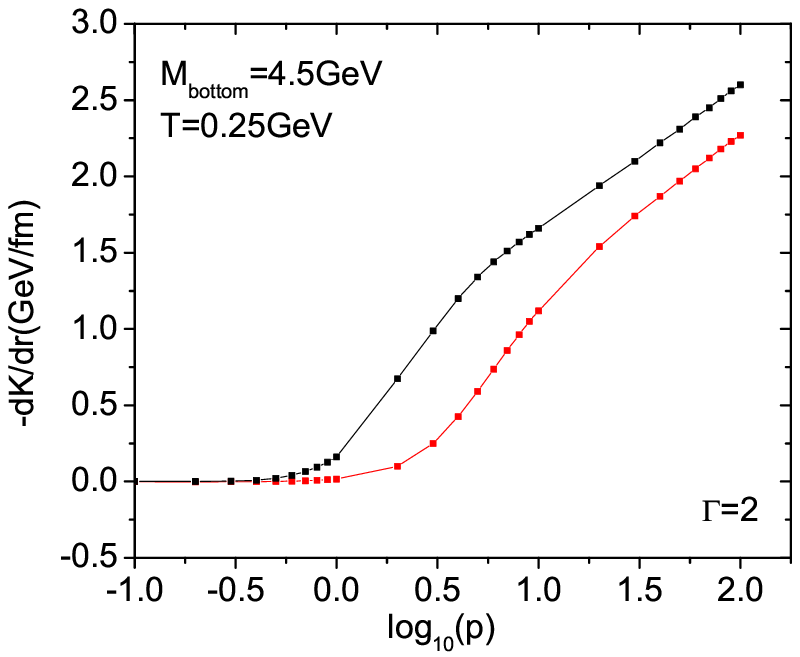}
\end{center}
\begin{center}
\includegraphics[width=0.55\textwidth]{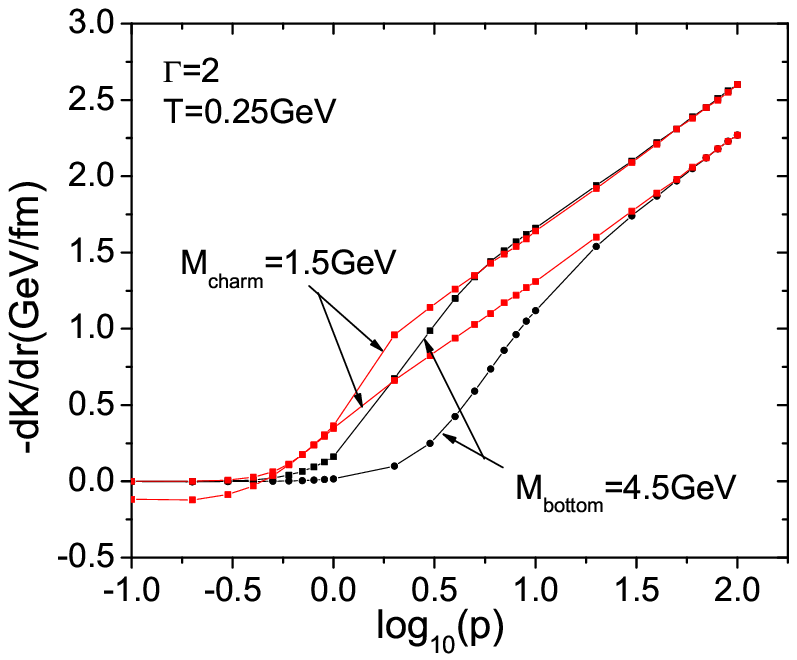}
\end{center}
\caption{Logarithmic energy loss for charm (left) and bottom (right) in absolute units. See text.}
\label{charm_bottom_log}
\end{figure}

\begin{eqnarray}
-\frac{dK}{dr}&=&3\Gamma^2\left(\frac{C_F}{C_2}\right)\frac{v_T^2}{v^2}\,\frac{T}{a_{WS}}
\int_0^{q_{max}}\,{dq}\,\frac{1}{q}\,\nonumber\\
&\times&\frac{1}{\sqrt{2\pi}}\int_{-v/v_T}^{v/v_T}\,dx\,e^{x^2/2}\,
\Bigg(\bigg((1-n{\bf c}_{D1}(q))\,e^{x^2/2}/x+n{\bf c}_{D1}(q)\,\psi(x)\bigg)^2+\pi\,n^2{\bf c}^2_{D1}(q)/2\Bigg)^{-1}\nonumber\\
\label{CB4}
\end{eqnarray}
where $q=k a_{WS}$ and $a_{WS}$ is the Wigner-Seitz radius.  The
units for the energy loss per length in (\ref{CB4}) follows from
$T/a_{WS}$. For SU(2), $C_F=3/4$ for a heavy quark, and $C_2=2$
for thermal constituent gluons of mass $m\approx \pi\,T$.
$a_{WS}=({3}/{4\pi
n})^{1/3}=(\frac{3}{4\pi}\frac{\beta^3}{0.244\times
3})^{1/3}=0.6883\beta$ for a density dominated by black-body
(gluon) radiation $n=0.244(N_c^2-1)/\beta^3=0.244\times
3/\beta^3$.

In Fig.~\ref{charm_bottom1} we show the dimensionless energy loss
following from (\ref{CB4}) for charm and bottom as a function of
the probe momentum $\gamma Mv$, for different $\Gamma=2,3,4$
around the SU(2) liquid point. The numerics have been carried
using the analytic structure factor of
Fig.~\ref{structure_G020304}. The energy loss is normalized to the
total kinetic energy in length $L$, $E/L=(\gamma-1)M/L$. Since the
quark velocity is maintained constant, the energy loss is seen to
exceed 1 for $\Gamma=4$. The loss is very sensitive to the Coulomb
coupling $\Gamma$ in the liquid phase.

In Fig.~\ref{charm_bottom_log} we show the energy loss
on a logarithmic momentum scale for both charm and bottom.
The upper curve (black) is the total loss from (\ref{CB4}),
while the lower curve (red)  is just the metallic loss following from (\ref{CB3}).
The difference is a measure of the energy loss due to collisions with the low
momentum part of the excitational spectrum of the SU(2) plasma which
is plasmon dominated. These are the wings shown in Fig.~\ref{2dplot}.
Charm and bottom jets with low momenta say $p\approx $ 3 GeV experience
energy loss through broad plasmons. The energy loss for jets with $p$ larger
than 10 GeV is mostly linear and therefore metallic.

\begin{figure}[!h]
\begin{center}
\includegraphics[width=0.49\textwidth]{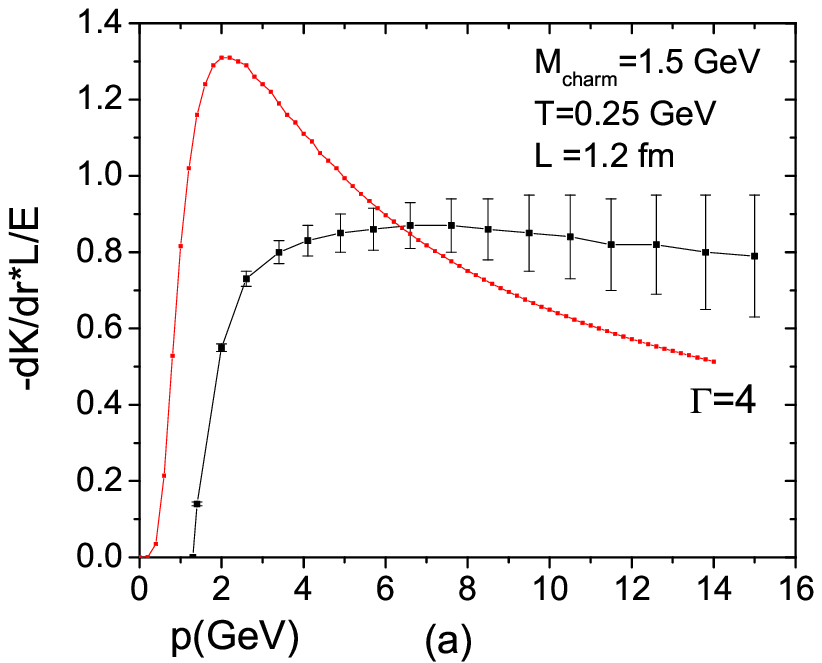}
\includegraphics[width=0.49\textwidth]{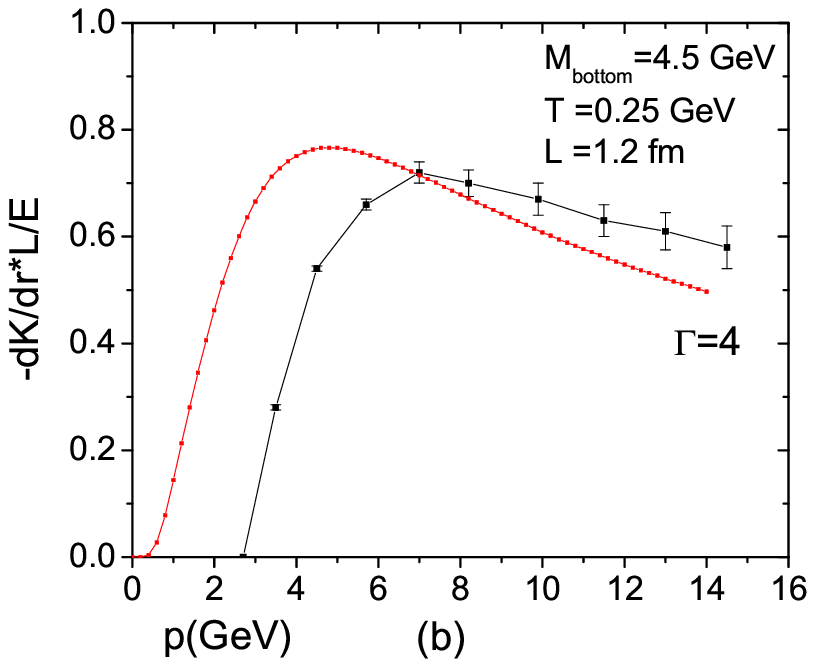}
\end{center}
\begin{center}
\includegraphics[width=0.49\textwidth]{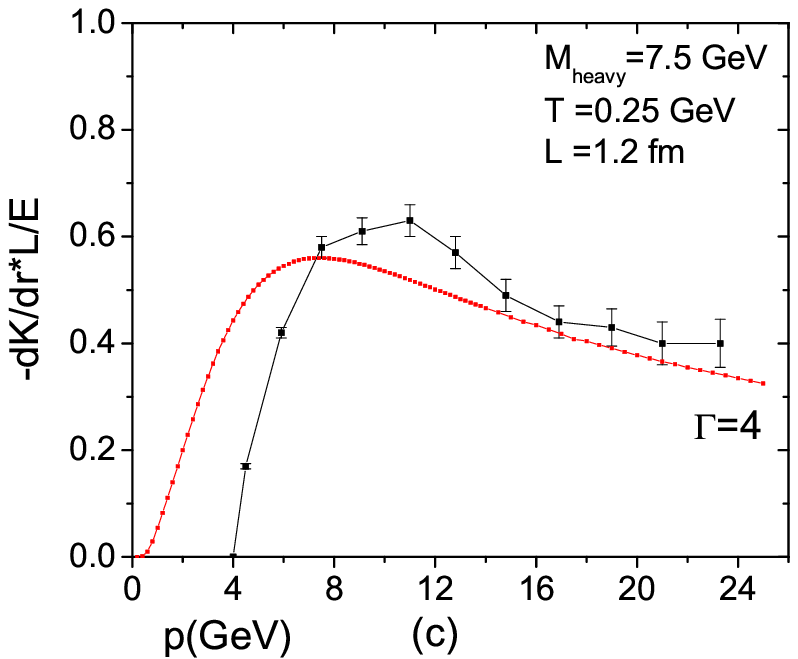}
\end{center}
\caption{Energy loss: (red) analytical versus (black)  SU(2)
molecular dynamics~\cite{dusling&zahed}.} \label{charm_bottom2}
\end{figure}

In Fig.~\ref{charm_bottom2} we compare our analytical results for
the energy loss (red curve) to recent SU(2) numerical simulations
(black curve) using the same model~\cite{dusling&zahed}.  We note
that the numerical simulations in~\cite{dusling&zahed} are quoted
for the mean-potential to kinetic energy ratio $V/K\approx 3$
which happens to fluctuate by about $\pm 1$ inside the simulation
box. To make the comparison meaningful, it is better to use in the
notation of~\cite{dusling&zahed} $\Gamma=1/Ta_{WS}\approx 4.8$ for
$n=1/\lambda^3$ and $\lambda=1/3T$ with $\lambda$ the minimum of
the potential in the same notations.  With this in mind, our
analytical results at $\Gamma=4$ compare favorably with the
molecular dynamics simulations at large momenta and for heavier
quark masses (say bottom). Most of  the discrepancy with the
simulations is at low momentum where the effects of the hard core
in~\cite{dusling&zahed} are the largest.

\section{Conclusions}

We have analyzed the energy loss of fast moving charm and bottom quarks in an SU(2)
color Coulomb plasma for a broad range of the Coulomb coupling.  The Coulomb character
of the underlying interaction retained classically make the energy loss entirely described by the
longitudinal part of the dielectric function. We have used linear response theory to derive
an explicit expression for  the imaginary part of the dielectric function in terms of the Laplace
transform of the time-dependent structure factor in the SU(2) Coulomb plasma.

We have shown that the probe initial color and statistical averaging causes the longitudinal dielectric
function to select the $l=1$ color channel of the time-dependent structure factor which is the plasmon
channel. The sound channel dominates the low momentum of the $l=0$ color channel, and decouples
from the longitudinal part of the dielectric function. While the SU(2) plasmon survives at strong coupling,
its width for $k>k_D$ is substantial and therefore causes it to thermally decay.

The energy loss of fast moving charm and bottom quarks is mostly due to the metallic aspect of the SU(2)
colored Coulomb plasma which is dominated by thermal particles.  There is no colored Cerenkov cone
as the plasmon is dwarfed in the metallic limit, nor a colorless Mach cone as the sound decouples due to
the probe initial colors. The energy loss is soundless. Our results are of course only classical. They apply for
a broad range of $\Gamma$ near the liquid point. The comparison to the MD simulations show that our
energy loss is about comparable at higher momenta and for heavier quarks where the effects of the numerical
hard core is small.  As initially reported in~\cite{dusling&zahed}, the energy loss is sizable.

Strong coupling assessment of jet energy loss in gauge theories
have been carried out in the context of holographic
QCD~\cite{herzogetal}.  The fact that a Mach cone was
reported in these calculations~\cite{chesler&yaffe},  maybe due to
the fact that the probe jet is actually colorless. Indeed, most of the
holographic jets are inserted with an external hand that maintains
a constant velocity and perhaps even  balance the color charge. Clearly
colorless (mesonic) jets of the type $\overline{Q}Q$  do couple to
the sound channel in our case through the ${\bf S}_{00}(k)$
structure factor~\cite{cho&zahed3}, and are expected to be trailed
by a Mach cone.

Finally, to carry our analysis of charm and bottom at RHIC and perhaps even LHC, require an assessment
of the heavy quark composition in the prompt phase of the heavy ion collision which we have not carried out.
Also, we need to address more carefully the correspondence between our classical SU(2) QGP and the
quantum SU(3) QGP. These issues will be addressed next.

\begin{acknowledgements}
We thank Kevin Dusling for discussions.
This work was supported in part by US DOE grants DE-FG02-88ER40388
and DE-FG03-97ER4014.
\end{acknowledgements}

\end{document}